\documentclass[apj,numberedappendix]{emulateapj}
\usepackage{apjfonts}

\shorttitle{Thermonuclear Supernovae from Ultracompact Binaries}
\shortauthors{Bildsten, Shen, Weinberg, and Nelemans}

\begin{document}

\title{Faint Thermonuclear Supernovae from AM Canum Venaticorum Binaries} 

\author{Lars Bildsten, Ken J. Shen}
\affil{Kavli Institute for Theoretical Physics and Department of Physics
Kohn Hall, University of California, Santa Barbara, CA 93106;\\
bildsten@kitp.ucsb.edu, kenshen@physics.ucsb.edu}

\author{Nevin N. Weinberg} 
\affil{Astronomy Department and Theoretical Astrophysics Center
601 Campbell Hall, University of California, Berkeley, CA 94720;
nweinberg@astro.berkeley.edu}

\author{Gijs Nelemans} 
\affil{Department of Astrophysics, Radboud University
Nijmegen, Toernooiveld 1, NL-6525 ED, The Netherlands;
nelemans@astro.ru.nl}

\begin{abstract}

Helium that accretes onto a Carbon/Oxygen white dwarf in the double white dwarf AM Canum Venaticorum (AM CVn) binaries undergoes unstable thermonuclear flashes when the orbital period is in the 3.5-25 minute range.  At the shortest orbital periods (and highest accretion rates, $\dot M>10^{-7} M_\odot \ {\rm yr^{-1}}$), the flashes are weak and likely lead to the Helium equivalent of classical nova outbursts. However, as the orbit widens and $\dot M$ drops, the mass required for the unstable ignition increases, leading to progressively more violent flashes up to a final flash with Helium shell mass $\approx 0.02-0.1 M_\odot$. The high pressures of these last flashes allow the burning to produce the radioactive elements $^{48}$Cr, $^{52}$Fe, and $^{56}$Ni that power a faint ($M_V=-15$ to $-18$) and rapidly rising (few days) thermonuclear supernova. Current galactic AM CVn space densities imply one such explosion every 5,000-15,000 years in $10^{11} M_\odot$ of old stars ($\approx 2-6$\% of the Type Ia rate in E/SO galaxies).  These ``.Ia''  supernovae (one-tenth as bright for one-tenth the time as a Type Ia supernovae) are excellent targets for deep (e.g. $V=24$) searches with nightly cadences, potentially yielding an all-sky rate of 1,000 per year.

\end{abstract}

\keywords{binaries: close--- 
novae, cataclysmic variables---
supernovae: general---
white dwarfs}

\section{Introduction}

Stellar evolution can lead to double white dwarf (WD) binaries, often in orbital periods ($P_{\rm orb}$) short enough that gravitational wave losses drive them into contact (Webbink 1984). Prior to making contact, these binaries are sources for space-based gravitational wave interferometers (Nelemans et al. 2001, 2004) and have been found in numerous optical surveys (Marsh et al. 1995; see Nelemans et al. 2005 for the latest overview).  For comparable mass WDs, the outcome at contact (when the lower mass WD fills its Roche lobe) is likely catastrophic (Webbink 1984; Marsh et al. 2004).

For extreme mass ratios (hence the lower mass WD is pure He; Marsh et al. 2004), the mass transfer at initial Roche lobe filling is dynamically stable (see Deloye \& Taam 2006 for a recent discussion). Under the continued loss of orbital angular momentum from gravitational radiation (Faulkner et al. 1972; Tutukov \& Yungelson 1979; Nelemans et al. 2001; Deloye, Bildsten \& Nelemans 2005), the binary separates to $P_{\rm orb}\approx 60$ minutes as the Helium donor gets whittled down to $< 0.01M_\odot$ in over $10^9$ years. This AM Canum Venaticorum (AM CVn) class of $P_{\rm orb}< 60$ minute binaries (Warner 1995; Nelemans 2005) has been known for 30 years, but recent harvesting of the Sloan Digital Sky Survey (SDSS, Roelofs et al. 2005; Anderson et al. 2005) and X-ray detections of possible direct impact binaries (Cropper et al. 1998; Israel et al. 2002; Ramsay et al. 2002; Marsh \& Steeghs 2002) have greatly increased the sample. The number of AM CVn binaries found in SDSS imply a space density of $(1-3)\times 10^{-6} \ {\rm pc^{-3}}$, including those which started mass transfer 10 Gyrs ago (Roelofs et al. 2007b)

We begin in \S \ref{sec:heflashes} by discussing the fate of the Helium that accretes on the WD at rates $10^{-9} - 10^{-6} M_\odot\ {\rm yr^{-1}}$. The burning is unstable for these accretion rates, leading to successive shell flashes likely to appear as classical novae. However, we find that the last unstable flash at low $\dot M$'s has a mass in the $0.02-0.1M_\odot$ range, large enough to become dynamical and eject radioactive $^{48}$Cr, $^{52}$Fe, and $^{56}$Ni. In \S 3, we show that the resulting thermonuclear supernovae are faint ($M_V\approx -16$) and rapidly evolving ``.Ia'' events. We close in \S 4 by deriving their occurrence rate in old stellar populations and discussing the prospects for their detection.

\section{Helium Shell Flashes in AM CVn Binaries} 
\label{sec:heflashes}

Before discussing the Helium shell flashes expected from AM CVn binaries, it is important to mention the original work on thicker helium shell flashes in the context of Helium star donors (i.e. stars actively burning helium with initial masses $M_d\approx 0.4M_\odot$). These are studied as another potential progenitor for thermonuclear supernovae via the double detonation scenario (Nomoto 1982a, 1982b; Woosley et al. 1986; Woosley \& Weaver 1994).  Once the helium star comes into contact (Iben \& Tutukov 1987, 1991), the mass transfer rate is set by gravitational wave losses (Savonije et al. 1986; Ergma \& Fedorova 1990), leading to the $\dot M-M_d$ trajectory (from Yungelson 2007, this binary started at 80 minutes and came into initial contact after some He burning had occurred) shown by the dotted line in Figure 1.  These donors adjust their thermal state until $P_{\rm orb}\approx 10$ minutes (where $M_d\approx 0.2M_\odot$), and then expand under further mass loss. Most of the Helium is added to the C/O WD at a rate of $2-6\times 10^{-8}M_\odot \ {\rm yr^{-1}}$.

The solid lines are  Iben \& Tutukov's (1989) estimated He ignition mass ($M_{\rm ign}$) dependence on $\dot M$ for different C/O WD masses. Calculated with the He burning star scenario in mind, these are  accurate for $\dot M< 10^{-7}M_\odot \ {\rm yr^{-1}}$, and compare well to numerical work (Nomoto 1982a; Fujimoto \& Sugimoto 1982;  Limongi \& Tornambe 1991; Yoon \& Langer 2004).  This illustrates that the large He ignition masses ($>0.2M_\odot$) needed to detonate the Helium and the underlying C/O  are achieved in this scenario for $0.6 M_\odot$ WDs. Yoon \& Langer (2004) found much lower $M_{\rm ign}$'s at these $\dot M$'s when they included large amounts of rotational dissipation at the depth of Helium burning. However, the work of Piro \& Bildsten (2004) showed that the accreted material comes into co-rotation with the star at much lower depths, so that there is no rotational impact on Helium ignition.

\begin{figure}
	\plotone{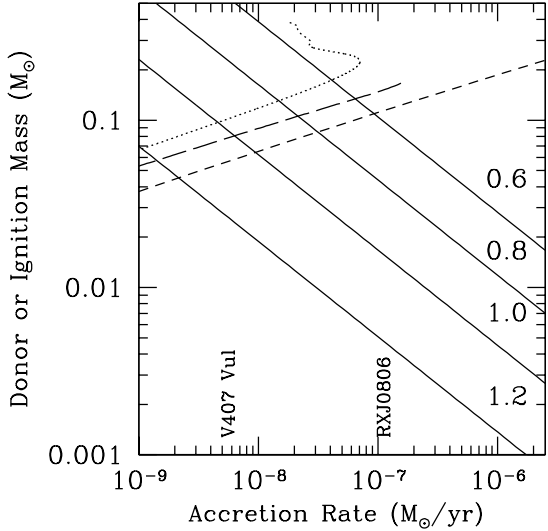}
	\caption{Fully degenerate (semi-degenerate) Helium donor mass in AM CVn binaries are shown by the \emph{dashed line} (\emph{long-dashed line} from C. Deloye, priv. comm.) as a function of $\dot M$, while the \emph{dotted line} is for a He burning  star (Yungelson 2007).   The \emph{solid lines} display Iben \& Tutukov's (1989) estimates of Helium ignition masses for pure He accretion on WDs with $M=0.6, 0.8, 1.0$ \& $1.2M_\odot$ as a function of $\dot M$.  The names of the two AM CVn binaries with $\dot{M}$'s in this range are shown at their respective $\dot{M}$'s.}
	\label{fig:heignmasses}
\end{figure}

Our focus here is the Helium ignition expected in the AM CVn binaries that contain a non-burning helium donor and a temporally declining $\dot M$.  The short-dashed line in Figure 1 shows the $\dot M-M_d$ relation (decreasing with time as the orbit widens) for a fully degenerate He donor. We begin the plot for $\dot M<2\times 10^{-6} M_\odot \ {\rm yr^{-1}}$, as the Helium burns stably at higher $\dot M$'s (e.g. for a $>0.2-0.27M_\odot$ He WD donor at $P_{\rm orb}\approx 2.5-3.5$ min; Tutukov \& Yungelson 1996), potentially appearing as a supersoft source (van Teeseling et al. 1997). However, as $\dot M$ decreases, the Helium burns unstably via recurrent flashes. For an accreting WD with $M=0.6M_\odot$, there will be $\approx 10$ recurrent weak He flashes as the donor mass drops from $0.2$ to $0.1M_\odot$.  Deloye et al. (2005) and Deloye \& Taam (2006) showed that a more realistic expectation is that the He donors have finite entropy (see also Roelofs et al. 2007a), which increases (\emph{long-dashed line}) the $\dot M$ for a fixed $M_d$, nearly to the dotted  line case, which is the helium burning star donor after adiabatic expansion quenches the burning.

In all these cases the ``last flash'' will have the largest $M_{\rm ign}$.  For simplicity, we use the intersection of $M_{\rm ign}$ with the degenerate donor mass in Figure 1 to find the fiducial last flash masses $M_{\rm ign}=0.11, \ 0.09, \ 0.06, \ 0.04 \ \&\ 0.02 M_\odot$ for $M=0.6, 0.8, 1.0, 1.2 \ \& \ 1.36 M_\odot$ WDs, shown as the short-dashed line in Figure 2.  Given the uncertainties in binary evolution, we allow for larger flash masses of 50\% (the increase if we find intersections with the dotted line in Figure 1) and lower flash masses of a factor of two (to accomodate the phase of the last flash and the possibility that $M_{\rm ign}$ may be lower than given by Iben \& Tutukov (1989)), resulting in the hatched region of Figure 2.
 
Once ignited, the flash proceeds by developing a convective zone that moves outward as the base temperature rises due to the energy generation rate from the triple-$\alpha$ reaction, $\epsilon_{\rm nuc}$ (Nomoto 1982a).  The thermonuclear runaway can lead to a dynamical burning event and potential detonation (Taam 1980a, 1980b; Nomoto 1982b) if the thermonuclear timescale, $t_{\rm nuc}=c_P T / \epsilon_{\rm nuc}$, becomes as short as the local dynamical time, $t_{\rm dyn} = H/c_s$, where $H=P/\rho g$ is the pressure scale height and $c_s$ is the sound speed.

In order to calculate these timescales, we integrate hydrostatic equilibrium for an isothermal C/O core (including Coulomb corrections to the ion pressure) and a fully convective pure He envelope.  Because $M_{\rm ign}$ can be a significant fraction of the core mass, the decreasing pressure during the convective evolution allows for hydrostatic expansion of the core, so we must calculate the structure of the core and envelope consistently. (Our calculation thus differs from that of Fujimoto \& Sugimoto (1982), who assumed that the core did not adjust during the flash.) For a given core mass and $M_{\rm ign}$, the radial expansion of the burning envelope leads to a maximum temperature that sets the minimum value of $t_{\rm nuc}/t_{\rm dyn} $. This minimum value decreases as $M_{\rm ign}$ is increased.  Thus, for each core mass, there is a minimum $M_{\rm ign}$ for a fully convective, hydrostatic envelope that can become dynamical during its expansion.

\begin{figure}
	\plotone{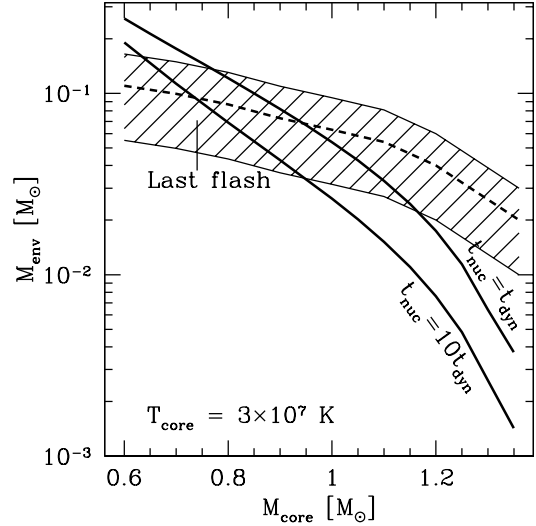} 
	\caption{Minimum $M_{\rm ign}$ needed to achieve $t_{\rm nuc}=t_{\rm dyn}$ and $t_{\rm nuc}=10t_{\rm dyn}$, as a function of the core mass (\emph{thick solid lines}).  An isothermal core of $3\times10^7$ K is assumed (Bildsten et al. 2006).  The \emph{dashed line} shows the ``last flash'' $M_{\rm ign}$, with an uncertainty of 50\% above and a factor of two below (\emph{hatched region}).}
	\label{fig:hemaxtemp}
\end{figure}

The thick solid lines in Figure 2 show the minimum $M_{\rm ign}$'s that achieve $ t_{\rm nuc}=t_{\rm dyn}$ and $ t_{\rm nuc}=10t_{\rm dyn}$
(to allow for the possibility of initiations in turbulent fluctuations; c.f. Wunsch \& Woosley 2004). Also shown is the estimated range of the ``last flash'' $M_{\rm ign}$ (\emph{hatched region}). For a C/O WD with $M>0.7-0.9 M_\odot$, the last flash is likely to be dynamical. For lower mass WDs, whether or not the last flash is dynamical will depend on the binary evolution and accurate calculations of $M_{\rm ign}$. However,  stable mass transfer at Roche lobe contact favors extreme mass ratios, potentially leading to more massive WDs in AM CVn's (Marsh et al. 2004), as recently inferred (e.g. Roelofs et al. 2007a).

These Helium flashes have ignition pressures lower ($P \approx 10^{22}-10^{23} \rm{ erg\  cm}^{-3}$) than previously studied in detail. Hashimoto et al. (1983) showed that constant pressure burning at these $P$'s can produce the radioactive elements $^{44}$Ti, $^{48}$Cr, $^{52}$Fe, and $^{56}$Ni. We ran a few detonation calculations using a one-dimensional, explicit, Lagrangian, finite difference scheme in which the hydrodynamics are coupled to a 13 isotope $\alpha$-chain reaction network. The network includes $\alpha$-chain, heavy-ion, and $(\alpha, p)(p, \gamma)$ reactions. Starting from the fully convective profile at the moment $t_{\rm nuc} = t_{\rm dyn}$, we perturb the temperature at the base of the envelope by 1\%.  A steady detonation forms on a dynamical time and propagates outward, steepening as it moves down the density gradient. A weak shock (over-pressure $\sim 2$) propagates downward into the C/O WD but does not ignite the Carbon at the envelope/core interface.  In general, we find products of He envelope burning similar to those found by Hashimoto et al. (1983). For example, for a $1\ M_\odot$ C/O WD with a $0.06M_\odot$ Helium envelope, at the time the shock reaches the WD surface,  $^{56}$Ni is the most abundant ash ($\approx 0.012M_\odot$), followed by $^{52}$Fe ($0.0076M_\odot$),  $^{48}$Cr ($0.0071M_\odot$), $^{44}$Ti ($0.0053M_\odot$), and unburned helium ($0.026M_\odot$).  Further burning will occur as the matter cools, but we have yet to calculate it fully.
 
We do not follow the shock deeper into the core of the WD, so the possibility remains for inwardly propagating shocks to detonate the Carbon at or near WD core (Livne 1990; Livne \& Glasner 1991; Woosley \& Weaver 1994; Livne \& Arnett 1995; Garc\'{i}a-Senz et al. 1999).  However, our range of last flash $M_{\rm ign}$'s is lower than every model that achieved a double detonation in the surveyed literature (Nomoto 1982b; Woosley et al. 1986; Livne 1990; Livne \& Glasner 1990, 1991; Woosley \& Weaver 1994), save for one data point in Nomoto (1982a), which is at the upper bound of our range of last flash $M_{\rm ign}$'s.  Note that Livne \& Arnett (1995) achieved double detonations inside our range of last flash $M_{\rm ign}$'s, but they artificially induced He detonations in models that should not have become dynamical.  Thus, it is likely that the last flash ignitions will not detonate the underlying WD, but this issue needs further examination.

\section{Radioactive Outcomes and Faint Supernovae} 
\label{sec:snovae} 

Given the diversity of outcomes, we only present preliminary calculations of the peak luminosity and time to peak light, $\tau_m$, when $M_e=0.02-0.1M_\odot$ of radioactive $^{56}$Ni (decay time $\tau=8.77$ d), $^{52}$Fe ($\tau=0.5$ d) or $^{48}$Cr ($\tau=1.3$ d) is ejected from the WD surface at the velocity $v\approx 15,000\ {\rm km \ s^{-1}}$ consistent with the nuclear energy release (minus the binding energy as it leaves behind the C/O WD).  The time of peak luminosity, $\tau_m$, is when the age since explosion matches the diffusion time through the envelope (for an opacity $\kappa$), yielding $\tau_m^2=\kappa M_e/7cv$ or 
\begin{equation}
	\tau_m\approx 2.5 {\rm d}\left(\kappa\over 0.1\  {\rm cm^2 \ g^{-1}}\right)^{1/2}
	\left(M_e\over 0.05M_\odot\right)^{1/2}
	\left(10^9 \ {\rm cm \  s^{-1}}\over v\right)^{1/2}
	\label{eq:arnett}
\end{equation}
where the factor 7 is found by both Arnett (1982) and Pinto \& Eastman (2000).  In Type Ia supernovae, the time to peak is 17-20 days (Conley et al. 2006; Garg et al. 2007), whereas the lower envelope masses in these explosions  give $\tau_m\sim {\rm days}$.  

Though the rapid rise of these events presents an observational challenge, our expectation is that they are very bright.  This is because from Arnett's (1982) ``rule''  we know that the peak luminosity matches that produced instantaneously by radioactive heating.  Hence the shorter $\tau_m$ allows us to see the power from the short-lived radioactivities, yielding peak bolometric luminosities of $1-5 \times 10^{42} \ {\rm erg \ s^{-1}}$, somewhat  lower than Type Ia supernovae (Contardo et al. 2000). These events come close to matching the peak brightnesses of sub-luminous Type Ia supernovae like 1991bg, though the observed long decay time requires an ejected mass  $> 0.1M_\odot$ (Stritzinger et al. 2006).

The radioactive heating from the $^{56}$Ni decay chain is well known, but the radioactive heating from $^{52}$Fe and $^{48}$Cr needs amplification.  The $^{52}$Fe decays (via positron emission 55\% of the time) with a half-life of 8.275 hours to the $1^+$ state of $^{52}$Mn, followed by a 168.7  keV $\gamma-$ray decay to the metastable $2^+$ state $^{52m}$Mn, giving an average heating of $0.86$ MeV.  The $^{52m}$Mn isotope has a 21.1 minute half-life and decays via positron emission to the $2^+$ state of $^{52}$Cr, releasing $3.415$ MeV.   The $^{48}$Cr isotope has a 21.56 hour half-life with nearly 100 \% electron captures to the $1^+$ excited state of $^{48}$V, followed by a cascade that emits an energy of $0.42$ MeV.  The $^{48}$V has a 15.973 day half-life, and an effective energy deposition of $2.874$ MeV. 

\begin{figure}
	\plotone{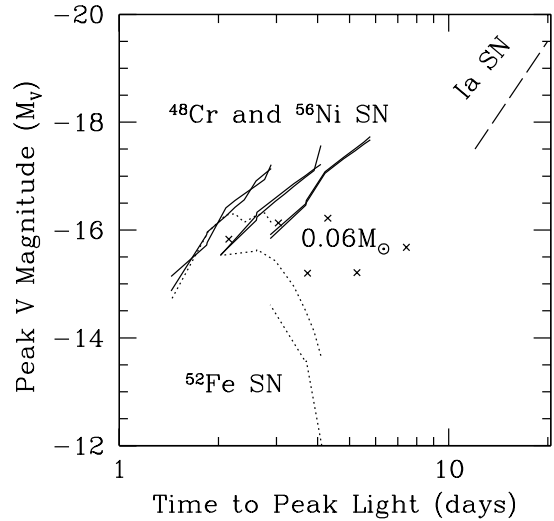}
	\caption{The peak magnitudes versus time to peak light of faint AM CVn supernovae powered by $^{48}$Cr/$^{56}$Ni (\emph{solid}) or $^{52}$Fe (\emph{dotted}) for (left to right) $\kappa=0.1,0.2,$ and $0.4 \ {\rm cm^2 \ g^{-1}}$.  Each line shows the range of ejected masses from $M_e=0.02-0.1M_\odot$ with velocities $v=12,000-17,000 \ {\rm km \ s^{-1}}$ (set by the underlying WD mass).  The crosses show the $0.06M_\odot$ explosion for the same three values of $\kappa$ and for $v\approx 15,700 \  {\rm km \ s^{-1}}$ (brighter set) and $v=\ 5,200 \ {\rm km \ s^{-1}} $ (dimmer set).  The \emph{long-dashed} line denotes the region occupied by Type Ia supernovae, including sub-luminous 1991bg-like events.}
	\label{fig:mvmax}
\end{figure}

In order to estimate the appearance of these events, we calculated $\tau_m$ from equation (\ref{eq:arnett}) with $\kappa$'s motivated by Type Ia SNe (Pinto \& Eastman 2000; Mazzali \& Podsiadlowski 2006) for the range of $M_e=M_{\rm ign}$ of the dashed line in Figure 2. For the velocities, we assumed that the matter leaves the WD with a kinetic energy equal to that available once the nuclear energy release unbinds the material.  In order to calculate the absolute V magnitude, $M_V$, shown in Figure 3, we assumed the radioactive heating luminosity  at $t=\tau_m$ for ejected envelopes of pure $^{56}$Ni  (\emph{solid curve}), $^{52}$Fe (\emph{dotted curve}), or $^{48}$Cr (\emph{solid curve}). By coincidence, the radioactive heating curves of $^{56}$Ni and $^{48}$Cr are similar at 2-10 days. For the brightest events ($^{48}$Cr and $^{56}$Ni), $T_{\rm eff}\approx 10,000-20,000$ K at maximum light (we assumed stellar colors to make our bolometric corrections), the ejected envelope is always radiation dominated, and the largest $M_{\rm ign}$'s are the brightest.  The rapid decay of $^{52}$Fe means that the largest $M_{\rm ign}$'s are the faintest (and have $T_{\rm eff}<3,000$K) since $\tau_m$ exceeds the $^{52}$Fe half-life. The low $M_{\rm ign}$ $^{52}$Fe events are brighter and have $T_{\rm eff}\approx 10,000$ K.  

Clearly work remains, including complete ignition and detonation calculations with  full nucleosynthetic outcomes and resulting velocity profiles.  In order to show what might be possible, the crosses detail the appearance of the $0.06M_\odot$ explosion reported in \S 2 for the same three values of $\kappa$, and  two values of $v$.  The lower value of $v$ was considered just in case some of the nuclear energy in the helium detonation is lost to shocking the underlying C/O WD. Clearly, even in that case, these remain bright events worthy of observational interest. 

\section{Discussion and Observational Consequences}
\label{sec:conclusion} 

Our consideration of the maximum Helium ignition masses naturally attained in AM CVn binaries led us to detonations that primarily  produce $^{48}$Cr, $^{52}$Fe, and $^{56}$Ni.  These rapidly decaying radioactive elements,  combined with the low ejected masses, cause rapidly rising (2-10 day) events with $M_V\approx -16$. Detection of such events would reveal the presence of AM CVn binaries in distant galaxies; all that remains is for us to estimate their rate. 

The measured local galactic density of AM CVn's (Roelofs et al. 2007b) yields a mass specific AM CVn birthrate in an old stellar population of  $(0.7-2)\approx 10^{-15}\  {\rm AM \ CVn} \ {\rm yr^{-1}} {L_{\odot, K}}^{-1}$.  If every AM CVn  gives a visibly explosive last flash, this would give 1 event every 5000-15000 years in a $10^{11}M_\odot$ E/S0 galaxy, $2-6$\% of the Type Ia SN rate in E/SO galaxies (Mannucci et al. 2005; Scannapieco \& Bildsten 2005; Sullivan et al. 2006). Since the time elapsed from reaching contact to the last flash is $<10^8$ years, the occurrence rate of these faint supernovae in an actively star forming galaxy depends on  the rate at which double WDs come into contact (Nelemans et al. 2004).  Additionally, if every AM CVn gives 20 weak Helium novae, then 1 in 5000-15000 classical novae would be a Helium-rich event.

For now, we use the local K-band luminosity density (Kochanek et al. 2001) to estimate the volume rate of .Ia's from the old stellar population to be $(0.4-1)\times 10^{-6} {\rm .Ia \ yr^{-1} \ Mpc^{-3}}$, $1.6-4$\% of the local Type Ia SN rate. Upcoming optical transient surveys (e.g. Pan-STARRS-1, Pan-STARRS-4, Large Synoptic Survey Telescope) with rapid cadences should find these events.  Observations with daily  cadences to a depth of $V=24$ would have an all-sky rate of up to $10^3$ yr$^{-1}$ (typical distance would be $\approx $ Gpc).  Since these .Ia's evolve quite rapidly, frequent followups, such as those provided by the Las Cumbres Observatory Global Telescope (www.lcogt.net) will prove a valuable resource for this endeavor. Their rapid evolution to small optical depths may also allow for strong $\gamma$-ray line emission.  

\acknowledgments
 
We thank D. Kasen and S. Kulkarni for discussions, J. Steinfadt for rate calculations, C. Stubbs for naming these events  .Ia's, and F. Timmes for providing the reaction network.  This work was supported by the NSF under grants PHY05-51164 and AST02-05956.  GN is supported by NWO Veni grant 639.041.405.

\end{document}